\documentstyle[sprocl]{article}
\input{psfig}
\newdimen\hhsize\hhsize=.5\hsize
\font\lloyd=cmr8 

\def\be{\begin{equation}}
\def\ee{\end{equation}}
\def\bea{\begin{eqnarray}}
\def\eea{\end{eqnarray}}

\def\be{\begin{equation}}
\def\ee{\end{equation}}
\def\bea{\begin{eqnarray}}
\def\eea{\end{eqnarray}}

\def\cmm2{{\,\rm cm^{-2}}}
\def\cm2{{\,{\rm cm}^2}}
\def\cmm3{{\,{\rm cm}^{-3}}}
\def\gcmm3{{\,{\rm g\,cm^{-3}}}}

\def\fun#1#2{\lower3.6pt\vbox{\baselineskip0pt\lineskip.9pt
  \ialign{$\mathsurround=0pt#1\hfil##\hfil$\crcr#2\crcr\sim\crcr}}}

\begin{document}
\title{
CMB POWER SPECTRUM ESTIMATION
}
\author{L. Knox, J.~R. Bond}
\address{Canadian Institute for Theoretical Astrophysics, 60 St.
George St., Toronto, ON M5S 3H8, CANADA}
\author{A.~H. Jaffe}
\address{Center for Particle Astrophysics, 301 LeConte Hall,
University of California, Berkeley, CA, 94720}
\maketitle\abstracts{
We explore power spectrum estimation in the context of a Gaussian
approximation to the likelihood function.  Using the Saskatoon
data, we estimate the power averaged through a set of ten filters
designed to make the errors on the power estimates uncorrelated.
We also present an improvement
to using the window function, $W_l$, for calculating bandpower estimates.
}

Estimates of parameters, $\theta_p$, from data will in general
have correlated errors, $\epsilon_p$; $C^{\rm P}_{pp'} \equiv 
\langle \epsilon_p \epsilon_{p'} \rangle$ is not necessarily diagonal.  
A Taylor expansion of the log of the likelhood function, $\ln{\cal L}$,
about the values of the parameter that maximize it, $\theta_p^*$, 
identifies $C^{\rm P}$ with the inverse of the second 
derivative of $\ln{\cal L}$.
The expectation value of this second derivative is an important quantity
known as the curvature matrix or Fisher matrix, $F$:

\bea
F_{pp'} \equiv -\langle{\partial^2 \ln{\cal L}(\theta_p) 
\over \partial \theta_p \partial \theta_{p'}}\rangle ={1\over 2}
{\rm Tr}\left[(C_T+C_n)^{-1}
{\partial C_T\over \partial \theta_{p}} (C_T+C_n)^{-1} 
{\partial C_T\over \partial \theta_{p'}} \right]
\eea
where $C_T$ and $C_n$ are the theory and noise covariance matrices,
$\langle \Delta \Delta^T \rangle = C_T + C_n$, where $\Delta$ is the
data (notation is more thoroughly explained in \cite{jkb}).  Since
${\partial^2 \ln{\cal L}(\theta_p) \over \partial \theta_p \partial
\theta_{p'}}$ is approximately equal to its expectation value, the
parameter covariance matrix, $C^P$, is approximately the inverse of the
Fisher matrix.  

Knowing $C^P$ is useful for two different purposes.  Firstly, it
is necessary if power spectrum estimation is to be used as a means of
``radical data compression''.  We have tried this with the Saskatoon
data\cite{nett} and the parametrization
${\cal C}_l = q_B {\cal C}_{l,B}^{\rm cdm}$, 
where ${\cal C}_l \equiv l(l+1)C_l/(2\pi)$ and 
${\cal C}_{l,B}^{\rm cdm}$  refers to standard, untilted cdm normalized
to $\sigma_8 = 1$ and restricted to $\ell$ within band $B$.
Having estimated $q_B$ for ten
contiguous evenly spaced bands 
from $\ell = 19$ to $\ell = 499$ we can approximate
the likelihood function of $\theta$, where $\theta$ is some other
parametrization of the spectrum:
\bea
-2\ln{\cal L}(\theta) \simeq \chi^2(\theta) = \left(q_A(\theta) - q_A\right)
F_{AB}\left(q_B(\theta)-q_B\right)
\eea
where $q_B(\theta) = 
\langle{\cal C}_l(\theta)\rangle_B/\langle{\cal C}_l^{\rm cdm}
\rangle_B$ and the brackets mean a logarithmic average across band $B$.   
See\cite{jkb} for how well this Gaussian approximation works when complicated 
slightly by a marginalization over calibration uncertainty.

The second use of $C^P$ (or equivalently the Fisher matrix) is for the
visual presentation of the power spectrum.  We can plot linear
combinations of power averaged through a set of filters where the
filters are designed to produce uncorrelated estimates.  One
particularly useful set of filters comes from Cholesky
decomposition\cite{Hamilton}, which is simply LU decomposition for a
symmetric matrix; find $L$ such that $F = LL^T$.  Notice now that
$L^{-1}$ does diagonalize $F$ since $L^{-1}F(L^{-1})^T$ is equal to
the identity matrix.  See $L$ in the left panel of the figure.  The
transformation affects the parameters by taking ${\bf q}$ to ${\bf Q}
= L^T {\bf q}$.  To convert the estimate of each $Q_\beta$ into a
bandpower estimate in band $\beta$, $\langle{\cal C}_l\rangle_\beta$
(plotted in the right panel of the figure), divide it by the sum over
the filter function, $f^\beta_B = L_{B\beta}/\langle{\cal C}^{\rm
cdm}_l\rangle_B$.  To find the bandpower prediction of another theory,
$\langle{\cal C}^{\rm t}_l\rangle_\beta$, average it over the filter
function: $\langle{\cal C}^{\rm t}_l\rangle_\beta = \sum_B f^\beta_B
\langle{\cal C}^{\rm t}_l\rangle_B/\sum_B f^\beta_B$.  Note that
$f^\beta_B$ is playing the role of $W_l/l$ in the usual bandpower
procedure\cite{bondbp}.  One can use $F^{1/2}$ instead of $L$ which
has been done with the COBE data\cite{mtkH}.

\centerline{\psfig{file=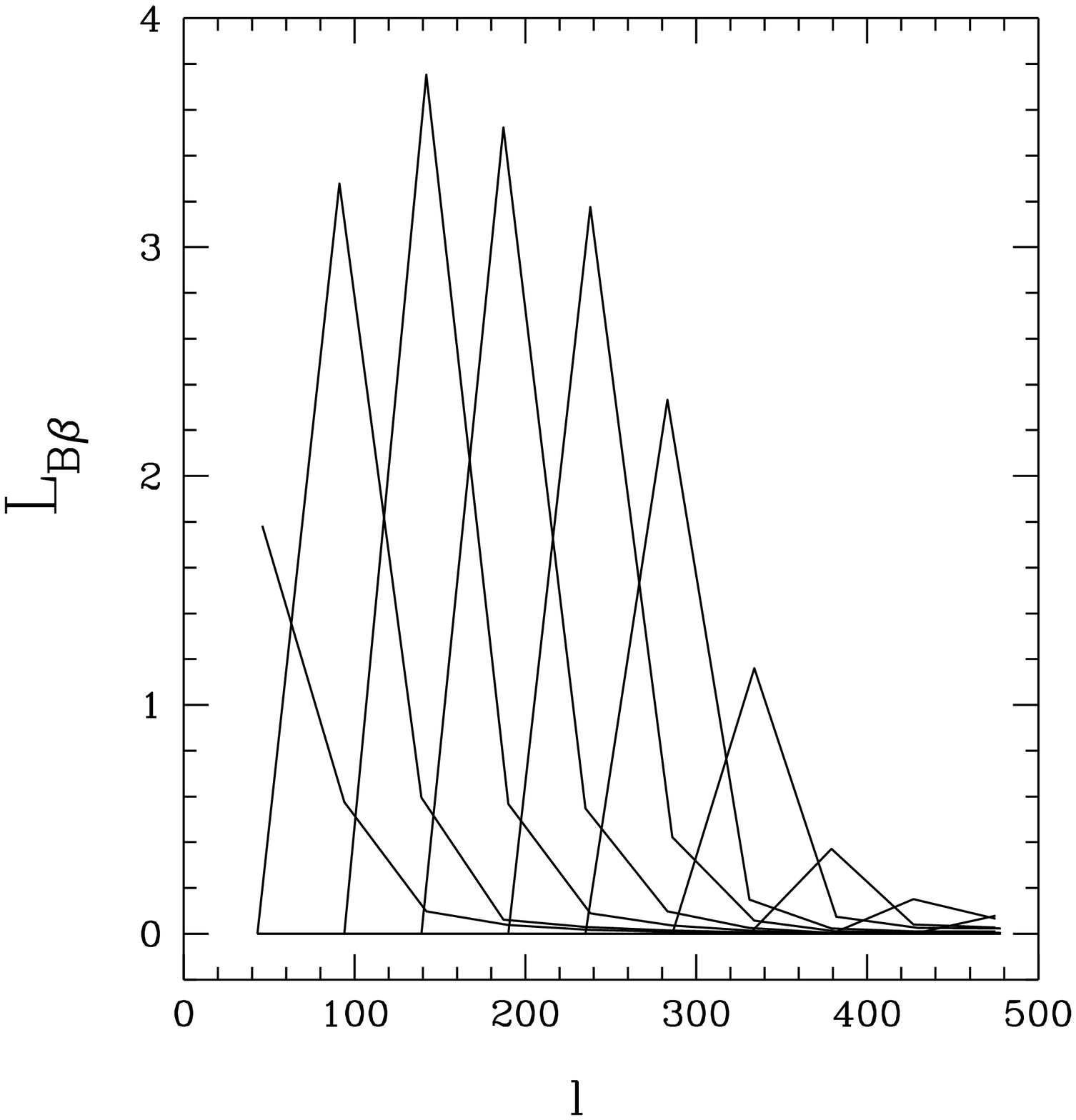,width=\hhsize,height=2.5in}
            \psfig{file=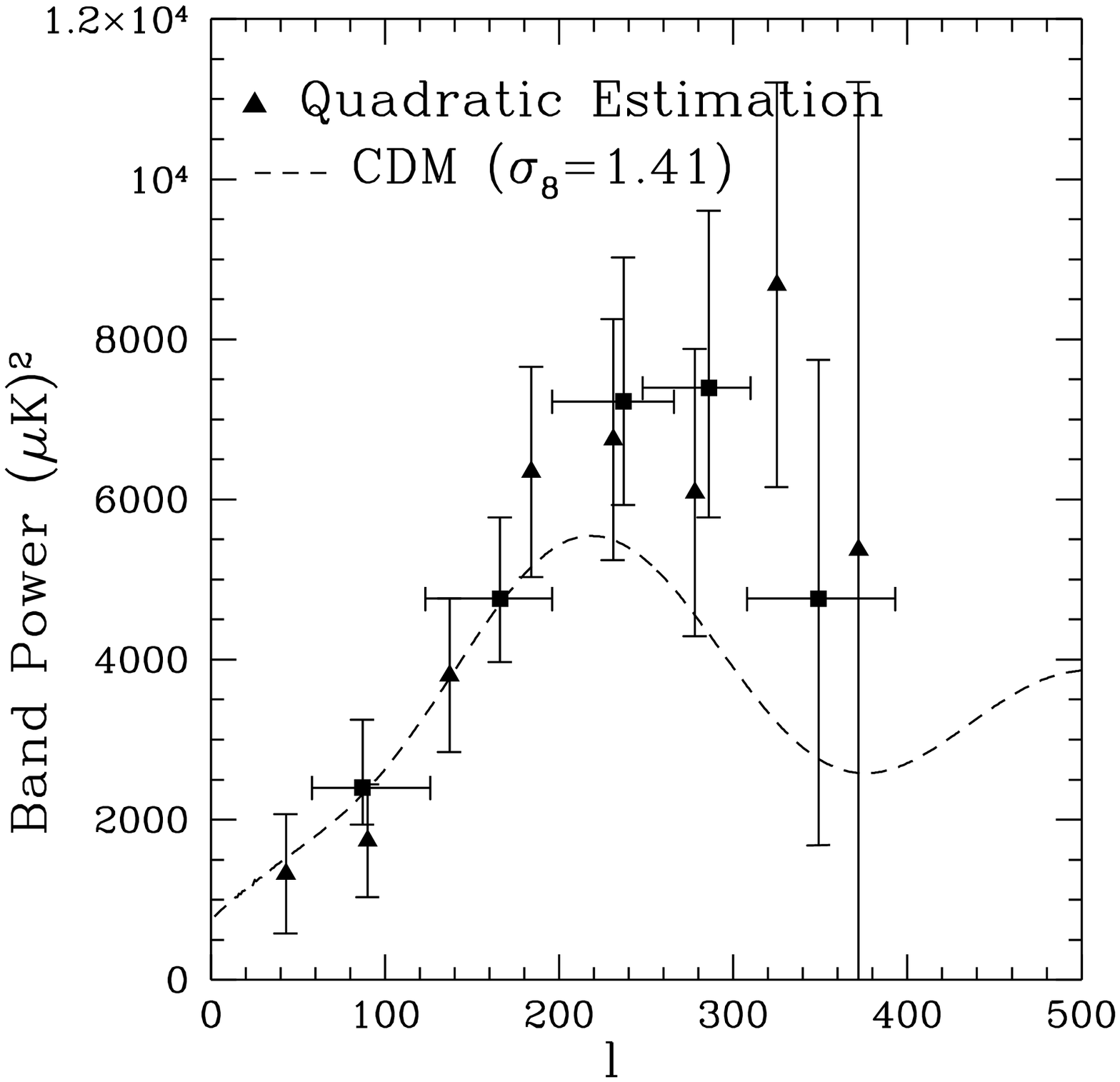,width=\hhsize,height=2.5in}}

\baselineskip=8pt 
{\lloyd 
\noindent Left panel: Cholesky decomposition of the Fisher matrix.
For each value of $\beta$ the
values of $L_{B\beta}$ at the $\ell$ value corresponding 
to the center of band $B$, are connected by straight lines. 
Right panel: Estimates of the power
spectrum from the SK data set as given by the observing team
(pentagons), and the quadratic estimator (triangles).  The 
quadratic estimates are uncorrelated because they are estimates
of power averaged over the filters $f^\beta_B$.      
The power estimates in the highest three bands have been averaged
together.}

\baselineskip=12pt

We estimated ${\bf q}$ with a quadratic estimator,
which can be derived from a Gaussian approximation to the likelihood
function, which the central limit theorem tells us is good in the
limit of large data sets.  The Gaussian approximation is equivalent to
truncating the Taylor series expansion of $\ln{\cal L}(\theta+\delta\theta)$
after the 2nd order term.  Doing so allows us to solve for $\delta\theta$
that maximizes the likelihood:
\bea
\delta \theta_p &=&
{1\over 2} (F^{-1})_{p p'}{\rm Tr}\left[\Delta \Delta^T (C_T+C_n)^{-1}
{\partial C_T\over \partial \theta_{p'}} (C_T+C_n)^{-1} - (C_T+C_n)^{-1}
{\partial C_T\over \partial \theta_{p'}} \right]\nonumber
\eea
Note that due to the matrix inversions this is an order of $n^3$
operation, where $n$ is the number of pixelized data points. 
Approximations to the weights $(C_T+C_n)^{-1}$ are necessary to
make this estimator practical for very large data sets.      

If we restrict ourselves to map-making experiments with no constraint
removals, and make the parameters the $C_l$'s then this reduces to the
quadratic estimator independently advocated by M. Tegmark \cite{mtk}.

The dependence of the right hand side on $\theta_p$ suggests an 
iterative approach.  The estimation of $\sigma_8$ for standard cdm took only 
three iterations starting
from $\sigma_8 = 1$ and converging on $\sigma_8 = 1.41 \pm 0.08$ 
(c.f. $1.43 \pm 0.08$ via the full likelihood analysis).  Therefore,
for the power estimates shown in the figure, we used the quadratic
estimator with $C_T$ for $\sigma_8 = 1.41$ standard cdm.



The bandpower expected from a given theory, ${\cal C}_l$, for a
given experiment with window function $W_l$ is usually calculated
by $(\sum_l {\cal C}_lW_l/l)/\sum_l W_l/l$
\cite{bondbp}.  The optimal (minimum variance) filter, however, is not
$W_l/l$ but instead $\sum_{l'}F_{ll'}$
where the parameters of this Fisher matrix are
${\cal C}_l$ and it is evaluated with values of ${\cal C}_l$ consistent
with the data and/or prior information.  Heuristically, this is an
inverse variance weighting.  We recommend that observers reporting
single bandpowers also report these filters which will
improve the bandpower method as a means of ``radical compression''.
In the limit that $C_T$ is proportional to the identity matrix the two filters are equivalent:  $\sum_{l'}F_{ll'} \propto W_l/l$.  

\section*{Acknowledgments} 
We would like to thank Andrew Hamilton for
a useful conversation.

\end{document}